\documentclass[sn-mathphys]{sn-jnl}

\usepackage{natbib}
\usepackage{epsfig}
\usepackage{subfig}
\usepackage{verbatim,amsmath}
\usepackage{threeparttable}
\usepackage{booktabs}

\renewcommand{\baselinestretch}{1.3}
\def\singlespace{\def\baselinestretch{1}\@normalsize}

\date{}

\renewcommand{\theequation} {\arabic{section}.\arabic{equation}}

\newcommand{\bY}{\mbox{\bf Y}}
\newcommand{\bZ}{{\bf Z}}
\newcommand{\bbeta}{\boldsymbol{\beta}}

\newcommand{\bSigma}{\boldsymbol{\Sigma}}
\def\wh{\widehat}
\newcommand{\rmT}{\rm T}

\jyear{2021}%

\theoremstyle{thmstyleone}%
\theoremstyle{thmstyletwo}%

\theoremstyle{thmstylethree}%

\begin{document}

\title[Model-free screening  for survival data based on HSIC]{Model-free screening procedure for ultrahigh-dimensional survival data based on Hilbert-Schmidt independence criterion}

\author[1]{\fnm{Xuerui} \sur{Li}}\email{xrli@bnu.edu.cn}

\author[2]{\fnm{Yanyan} \sur{Liu}}\email{liuyy@whu.edu.cn}

\author[2]{\fnm{Yankai} \sur{Peng}}\email{pengyk@whu.edu.cn}

\author*[3]{\fnm{Jing} \sur{Zhang}}\email{jing66@zuel.edu.cn}

\affil[1]{\orgdiv{Center for Statistics and Data Science}, \orgname{Beijing Normal University}, \country{Zhuhai 516087, China}}

\affil[2]{\orgdiv{School of Mathematics and Statistics}, \orgname{Wuhan University}, \country{Wuhan, Hubei 430072, China}}

\affil[3]{\orgdiv{School of Statistics and Mathematics}, \orgname{Zhongnan University of Economics and Law}, \country{Wuhan, Hubei 430073, China}}

\abstract{How to select the active variables which have significant
impact on the event of interest is a very important and
meaningful problem in the statistical analysis of ultrahigh-dimensional data.
Sure independent screening procedure has been demonstrated to be an effective
method to reduce the dimensionality of data from a large scale to a
relatively moderate scale.
For censored survival data,
the existing screening methods mainly adopt the
Kaplan--Meier estimator to handle censoring,
which may not perform well for scenarios which have heavy censoring rate.
In this article, we propose a  model-free screening procedure based on
the Hilbert-Schmidt independence criterion (HSIC).
The proposed method avoids
the complication to specify an actual model from a large number of covariates.
Compared with existing screening procedures,
this new approach has several advantages.
First, it does not involve the Kaplan--Meier estimator,
thus its performance is much more robust for the cases with a heavy censoring rate.
Second, the empirical estimate of HSIC is very simple as it  just depends on  the trace of a product of Gram matrices.
In addition,
the proposed procedure
does not require any complicated numerical optimization,
so the corresponding calculation is very simple and fast.
Finally, the proposed procedure which employs the kernel method is substantially
more resistant to outliers.
Extensive simulation studies demonstrate that
the proposed method has favorable exhibition over the existing methods.
As an illustration, we apply the proposed method to
analyze the diffuse large-B-cell lymphoma (DLBCL) data  and the ovarian cancer data.
}



\maketitle

\section{Introduction}
With the rapid advance of technologies,
ultrahigh-dimensional data are frequently encountered
in various fields of scientific
research, including genomics, biomedical imaging and economics.
In statistical analysis  of such data,
variable selection is an important issue
and many methods for it have been developed,
such as forward selection, backward selection, and best subset
selection.
Among them, the penalized estimation procedures,
such as Lasso (\citealp{Tibshirani1996}),
SCAD (\citealp{Fan2001}), MCP (\citealp{Zhang2010}),
have become increasingly popular
and  perform well in
variable selection and parameter estimation.
However, these regularization approaches may
face the challenges of
computational expediency, statistical accuracy and
algorithmic stability
when the dimension $p$ is ultrahigh in the sense
that $p=\exp(n^\alpha)$ with $\alpha>0$ (\citealp{Fan2009}).
To tackle these difficulties,
\citet{Fan2008} proposed the sure independence screening procedure (SIS)
in the context of a linear regression model
based on the marginal correlation between each covariate and the response variable.
They further proved that the SIS procedure can effectively reduce the dimension
while retaining all important covariates.
Following them, many authors extended the SIS procedure
to different models,
including the generalized linear model (\citealp{Fan2010}),
additive model (\citealp{Fan2011}),
the varying coefficient model
(\citealp{Fan2014,Liu2014}).
To avoid the complication to specify an actual model from a large number of covariates,
model-free screening procedures based on various marginal utilities have been proposed
(e.g., \citealp{Zhu2011,Li2012,Cui2015,Mai2015,Wu2015}).

In medical studies and clinical trials,
the outcome of interest is often the survival time $T$ of patients,
which may not be fully observed due to various reasons,
such as loss of contact with the patients,
individuals still alive at the end of the study, etc.
How to select the active variables which have a  significant
impact on $T$  among the huge number of covariates
is much more difficult due to the censoring.
Many researchers have studied this problem and proposed
several model-based screening methods
(e.g., \citealp{Tibshirani2009,Zhao2012,Gorst2013})
and model-free screening methods
(e.g.,
\citealp{Song2014,Zhang2017,Zhou2017,Liu2018,Zhang2018,Pan2019})
via defining different marginal utilities.
Most of these marginal screening methods
adopted the Kaplan--Meier estimator to handle censoring,
and have been proved to performs well in
in reducing the dimensionality for
ultrahigh-dimensional survival data.
However, these methods
may not perform well for scenarios which have a heavy censoring rate.
\citet{Zhang2021} proposed a model-free screening method based on
distance correlation, which avoided using the Kaplan--Meier estimator,
and it is much more robust than the other existing methods.

In this paper, we proposed a new model-free screening method for
ultrahigh dimensional right-censored  survival data.
Our method is based on a new independence  criterion named Hilbert-Schmidt independence criterion (HSIC), which is proposed by Gretton et al.(2005).
They proved that HSIC is indeed a dependence criterion under
all circumstances, i.e., HSIC is zero if and only if the random variables are independent.
This excellent property motivates us to use HSIC to filter out  some irrelevant covariates.
Compared with previous screening procedures,
this new approach has several distinctive advantages.
First, it does not involve the Kaplan--Meier estimator,
thus its performance is much more robust for the cases with heavy censoring rate.
Second, HSIC is  a measure of independence rather than correlation,
HSIC is zero means that the random variables are independent.
Moreover,
the empirical estimate of HSIC is very simple as it
is just the trace of a product of Gram matrices.
Third,
the proposed procedure which employs the kernel method is substantially
more resistant to outliers.
Finally, the proposed procedure
does not require any complicated numerical optimization,
thus the corresponding calculation is very simple and fast.
All these advantages greatly facilitate its
implementation in real applications.

The remainder of the article is organized as follows.
In Section 2,
we first introduce some notations and the definition of HSIC,
then present
the proposed model-free HSIC-based screening procedure in details.
Section 3 presents the results obtained from simulation studies.
Section 4 applies the proposed procedure to two real data sets.
Some discussions and concluding remarks are provided in Section 5.

\section{Proposed Method}
\subsection{Hilbert-Schmidt independence criterion}
Let $T$ denote the event time of interest,
$C$ denote the censoring time,
$\bZ=(Z_{1},\ldots,Z_{p})^{\rmT}$ denote the $p$-dimensional vector of covariates.
Furthermore, define $X=\min(T,C)$ and $\Delta=I(T\le C)$,
where $I(\cdot)$ denotes the indicator function.
Throughout this article, it is assumed
that $T$ and $C$ are conditionally independent given $\bZ$.
Consider a failure time study that consists of $n$ independent subjects,
the observed data can be summarized as
$\{X_i, \Delta_i, \bZ_i\equiv(Z_{1i},\ldots,Z_{pi})^{\rmT}: i=1,\ldots,n\}$,
which are independent copies of $(X, \Delta, \bZ)$.

In ultrahigh-dimensional scenarios, the number of covariates $p$ is much
larger than sample size $n$, and can grow exponentially with $n$,
e.g., $p=\exp(n^\alpha)$ with $\alpha>0$.
Under the sparsity principle,
only a small number of covariates have great influence on  $T$.
By following \citet{Song2014} and others,
the index set of these active covariates can be defined as
$$
\mathcal{A}=\{k: S(t\lvert\bZ){\text{functionally depends on}}\,\,Z_k, k=1,{\ldots},p\},
$$
where $S(t\lvert\bZ)=P(T>t\lvert\bZ)$ denotes the conditional survival function of $T$ given $\bZ$.
The goal in this paper is to estimate the active set $\mathcal{A}$ as precisely as possible based on the observed data.

Since the proposed screening utility in this paper is built based on Hilbert-Schmidt independence criterion (HSIC),
we first introduce its definition briefly.
Let $u$ and $v$ denote two random variables,
$\mathcal{U}$ and $\mathcal{V}$ denote separable space,
$\bf{\Gamma}$ and $\bf{\Lambda}$ being the Borel sets on
$\mathcal{U}$ and $\mathcal{V}$,
$p_{u,v}$ denote the joint measure on $(\mathcal{U}\times\mathcal{V},\bf{\Gamma}\times\bf{\Lambda})$,
$\mathcal{F}$ and $\mathcal{G}$ are the reproducing
kernel Hilbert space (RKHS) induced by $\mathcal{U}$ and $\mathcal{V}$,
$C_{uv}$ denote the cross-covariance operator associated
with the joint measure $p_{u,v}$ on $(\mathcal{U}\times\mathcal{V},\bf{\Gamma}\times\bf{\Lambda})$.
Following Definition 1 and Lemma 1 in \citet{Gretton2005},
the HSIC between  $u$ and $v$ is defined as
the sum of the squared singular values of the cross-covariance operator $C_{uv}$
(i.e., the squared HS-norm of $C_{uv}$),
which can be expressed as
\begin{eqnarray}\label{1}
 {\rm{HSIC}}(u,v)&=& {\rm{HSIC}}(p_{u,v},\mathcal{F},\mathcal{G})\notag \\
   &=& \| C_{uv}\|_{HS}^2\notag \\
   &=& E_{u,u',v,v'}[k(u,u')l(v,v')] +E_{u,u'}[k(u,u')]E_{v,v'}[l(v,v')]\notag \\
   &&-
   2E_{u,v}[E_{u'}[k(u,u')]E_{v'}[l(v,v')]],
\end{eqnarray}
where $\| \cdot\|_{HS}$ denotes the Hilbert-Schmidt (HS) norm,
$E_{u,u',v,v'}$ denotes the expectation over independent pairs $(u,v)$ and $(u',v')$
drawn from $p_{u,v}$,
$k$ and $l$ are unique positive definite kernel.
In Theorem 4 of \citet{Gretton2005},
they proved that
$ {\rm{HSIC}}(u,v)=0$ if and only if $u$ and $v$ are independent.
This excellent property motivates us using HSIC to filter out some irrelevant covariates
in our proposed screening procedure.
Given the observed data $\{u_i,v_i: i=1,2,\ldots,n\}$,
the estimator of ${\rm{HSIC}}(u,v)$ is given by
\begin{eqnarray}\label{2}
  \wh{\rm{HSIC}}(u,v) &=& (n-1)^{-2}tr(KHLH)
\end{eqnarray}
where $tr(\cdot)$ represents the trace of a matrix,
$H,K,L\in R^{n\times n}$,
$K$ and $L$ denote the corresponding kernel matrix of $u$ and $v$,
i.e.,
$K_{i,j}=k(u_i,u_j)$, $L_{i,j}=l(v_i,v_j)$,
$H_{i,j}=\sigma_{ij}-n^{-1}$,
$\sigma_{ij}=1$ for $i=j$ and 0 otherwise.
Compared with other kernel independence measures,
HSIC is simpler to define,
requires no regularization or tuning beyond kernel selection,
it is much more robust to outliers,
and
the finite sample bias of the estimate is negligible
compared to the finite sample fluctuations.
All these advantages greatly facilitate its
implementation in the proposed method.
For more details on the
definition of HSIC, please refer to \citet{Gretton2005}.

\subsection{HSIC-based screening procedure}
In this paper, the goal is to estimate the active set $\mathcal{A}$,
i.e., we want to find the active covariates which have great influence on  $T$.
As discussed above, to recover $\mathcal{A}$,
a natural screening utility is
the HSIC between $Z_k$ $(k=1,2,\ldots, p)$ and $T$, denoted as ${\rm{HSIC}}(Z_k,T)$.
${\rm{HSIC}}(Z_k,T)\neq 0$ means that
the $k$th covariate $Z_k$  is  an active covariate,
otherwise, $Z_k$  is  an inactive covariate.
However, for right-censored data,
${\rm{HSIC}}(Z_k,T)$ can not be estimated directly since
$T$ can not be fully observed,
so we set $(X,\Delta)$ as the two-dimensional response vector
and consider the HSIC between $(X,\Delta)$
and each covariate $Z_k~(k=1,\ldots, p)$.
Similar to \citet{Zhang2021},
we first standardize $(X,\Delta)$ marginally and define
$$X^*=\frac{X-\mu_{X}}{\sigma_{X}},~~ \Delta^*=\frac{\Delta-\mu_{\Delta}}{\sigma_{\Delta}},$$
where $\mu_{X}=\rm{E}(X)$, $\mu_{\Delta}=\rm{E}(\Delta)$, $\sigma_{X}=\rm{sd}(X)$, $\sigma_{\Delta}=\rm{sd}(\Delta)$,
$\rm{E}(\cdot)$ and $\rm{sd}(\cdot)$ represent
the expectation and standard deviation.
Then we set the standardized outcome as $\bY= (X^*,\Delta^*)$,
and define the marginal utility for feature screening as
\begin{equation}\label{3}
\omega_k={\rm{HSIC}}(Z_k,\bY),
\end{equation}
where ${\rm{HSIC}}(Z_k,\bY)$ denotes the HSIC between
$\bY$ and $Z_k$,
and is defined in the same way as equation (\ref{1}).
Following Theorem 4 in \citet{Gretton2005},
$\omega_k$ can serve as the population quantity of the proposed procedure for ranking the dependence between the covariate $Z_k$ and the failure time $T$.

Given the observed data
$\{X_i, \Delta_i, {\bf Z}_i \equiv(Z_{1i},\ldots, Z_{pi})^{\rmT}: i=1,\ldots,n\}$,
we can easily obtain the estimator $\wh{\bY}_i=(\wh{X}^*_i,\wh{\Delta}^*_i)$ ($i=1,\ldots,n$),
where
$$\wh{X}^*_i=\frac{X_i-\wh{\mu}_{X}}{\wh{\sigma}_{X}},~~\wh{\Delta}^*_i=\frac{\Delta_i-\wh{\mu}_{\Delta}}{\wh{\sigma}_{\Delta}},$$
$(\wh{\mu}_{X}, \wh{\sigma}_{X})$ and $(\wh{\mu}_{\Delta}, \wh{\sigma}_{\Delta})$ denote the sample mean and sample standard
error of the $X_i$'s and $\Delta_i$'s, respectively.
Following equation (\ref{2}),
we can obtain the estimator of $\omega_k$ as
\begin{equation}\label{3}
\wh{\omega}_k={\wh{\rm{HSIC}}}(Z_k,\wh{\bY}).
\end{equation}
Based on the discussion above and the property of HSIC (\citealp{Gretton2005}),
$\wh{\omega}_k$ is expected to fluctuate around zero if $Z_k$ is an inactive covariate,
and to be away from zero otherwise.
Then we can select  those candidate covariates with top values of $\wh{\omega}_k$
as active covariates.
This motivates the estimator of $\mathcal{A}$  given by
\begin{equation*}
\wh{\mathcal{A}}=\{j : \wh{\omega}_j \; \textrm{is amongst the first}\;
d_n \;\textrm{largest of all}\; \wh{\omega}_k, \, \ k = 1 , \ldots , p  \},
\end{equation*}
where $d_n$ is a pre-determined positive integer and suggested to be
$d_n = \lceil n/\log n\rceil$
(e.g., \citealp{Fan2008,Zhu2011,Li2012,Song2014}).
For brevity,
we refer to this HSIC-based screening procedure as HSIC-SIS.

\section{Simulation Studies}
In this section,
we evaluate the finite-sample performance of the proposed
screening procedure HSIC-SIS
and further compare it with the existing competitors.
For brevity, we refer to
the feature aberration at survival times screening procedure of
\citet{Gorst2013} as FAST-SIS,
the censored rank independence screening method of \citet{Song2014}  as CRIS,
the censored cumulative residual screening procedure proposed by  \citet{Zhang2018} as CCRIS,
the distance correlation-based screening method of \citet{Zhang2021} as DC-SIS.
Following \citet{Li2012}, we consider
the following criteria to measure the performance of different methods.
\begin{itemize}
\item [(i)] $\cal S$: the minimum model size required to include all active variables.
We present the median and interquartile range (IQR) of $\cal S$ over 200 replications.
\item [(ii)] ${\cal {P}}_e$: the selection proportion that each
active variable is selected into the model with the model
size  $d_n=\lceil n/\log n\rceil$ among all replications,
where $\lceil x\rceil $ denotes the integer part of $x$.
\item [(iii)] ${\cal {P}}_a$: the selection proportion that all
active variables are selected into the model with the model size $d_n$ as above.
\end{itemize}
An effective screening procedure is expected to yield $\cal S$ close
to the true minimum model size and both ${\cal {P}}_e$ and ${\cal P}_a$ close to one.

We generate the failure time  $T$ from
the Cox proportional hazards model, the nonlinear model,
and the general transformation model as described below.
Under each setting,
we consider two censoring mechanisms,
i.e.,
the completely random censoring mechanism
where ${C}\sim \rm{Unif}(0,\tau)$,
and
the informative censoring mechanism where $C\sim\rm{Unif}(0,c\cdot\lvert Z_1-Z_2\rvert)$,
$\tau$ and $c$
are chosen to yield the desirable censoring rates.
The results given below are based on  sample size $n=200$,
the number of covariates $p=2000, 5000, 10000$,
and the censoring rate CR=$20\%, 40\%$ with 200 replications.
In the proposed screening method HSIC-SIS,
the calculation of  $\wh{\omega}_k$ depends on  the selection of kernel function,
we consider the Gaussian kernel with $f(x,y)=\exp(-\frac{\|x-y\|^2}{2\gamma^2})$
with $\gamma=2$.

\textit{Example 1.}
We generated survival time
$T_i$ from the Cox proportional hazards model with the conditional hazard function given by
$$\lambda(t \lvert\bZ_i)=\lambda_{0}(t)\exp(\bZ_i^{\rm T}\bbeta),$$
where the baseline hazard function is set to be $\lambda_{0}(t)=(t-0.5)^2$ and
the ultrahigh-dimensional covariate $\bZ_i=(Z_{1i},\ldots,Z_{pi})^{\rm T}$ follows
a multivariate normal distribution with
 mean $\bf 0$ and correlation matrix
 $\bSigma=(0.8^{\lvert i-j\rvert})$ for $i,j=1,\ldots,p$.
 We set the true parameters $\bbeta=(0.35,0.35, 0.35, 0.35,$ $ 0.35, 0,
 \ldots, 0)^{\rmT}$, i.e., only the first five covariates
 $Z_1, Z_2,\ldots, Z_5$ are active covariate.
In this example, the index set of active covariates $\mathcal{A}=\{1,2,3,4,5\}$.
The simulation results under this model setting are summarized in Tables \ref{ex1.1}  and \ref{ex1.2},
from which we can see that the performance of the proposed HSIC-SIS procedure is
comparable with the model-free screening method DC-SIS and the
model-based screening method FAST-SIS,
both of them  outperform the model-free screening methods  CCRIS and CRIS.
In general, the proposed method HSIC-SIS performs well
for all the cases considered here,
even for the ultra-high dimensionality $p=10000$ or high censoring rate of $40\%$
or the informative censoring mechanism.

\textit{Example 2.}
To examine the performance of the proposed screening procedure for censored
nonlinear survival models with interactions,
we generated the log survival times from the model
$$\log T=(2+\sin Z_1)^2+(1+Z_5)^3+3Z_{10}^2+Z_1Z_{10}+\epsilon,$$
where $\epsilon\sim N(0,1)$,
$\bZ\sim N_p({{0}},\bSigma)$ with  $\bSigma=(0.8^{\lvert i-j\rvert})$ for $i,j=1,\ldots,p$.
In this example, only covariates $Z_1, Z_5, Z_{10}$ are active covariates,
i.e., the index set of active covariates $\mathcal{A}=\{1,5,10\}$.
Tables \ref{ex2.1}  and \ref{ex2.2} summarize the simulation results
for different procedures,
we can see clearly that the proposed screening method HSIC-SIS
is able to capture the nonlinear covariate effects with interactions.
The performance of the HSIC-SIS and DC-SIS procedures are comparable,
both of them perform significantly better than FAST-SIS, CCRIS and CRIS.

\textit{Example 3.}
We adopt the model setting of Song  et al. (2014) and
generate the survival time from the following transformation model
\begin{equation*}
H(T)=-\bbeta^{\rm T}\bZ+\epsilon,
\end{equation*}
where $\epsilon\sim N(0,1)$, $H(t)=\log\{0.5 (e^{2t}-1)\}$,
$\bZ\sim N_p({{0}},\bSigma)$ with
$\bSigma=(0.5^{\lvert i-j\rvert})$ for $i,j=1,\ldots,p$.
We set $\bbeta=(-1,-0.9,{\bf0}_6,0.8,1.0,{\bf0}_{p-10})^{\rm T}$,
i.e., only $Z_1, Z_2, Z_{9}, Z_{10}$ are active covariates,
the index set of active covariates $\mathcal{A}=\{1,2,9,10\}$.
The results are summarized in Tables \ref{ex3.1}  and \ref{ex3.2},
from which we can see that
the performance of the proposed
HSIC-SIS procedure is comparable with the mode-based method FAST-SIS and
model-free method DC-SIS,
furthermore, HSIC-SIS procudure exhibits more satisfactory
results than the model-free methods CCRIS and CRIS,
especially for cases with high censoring rate or  high dimensionality.

\begin{figure}[h]
	\centering
	\includegraphics[width=0.6\textheight]{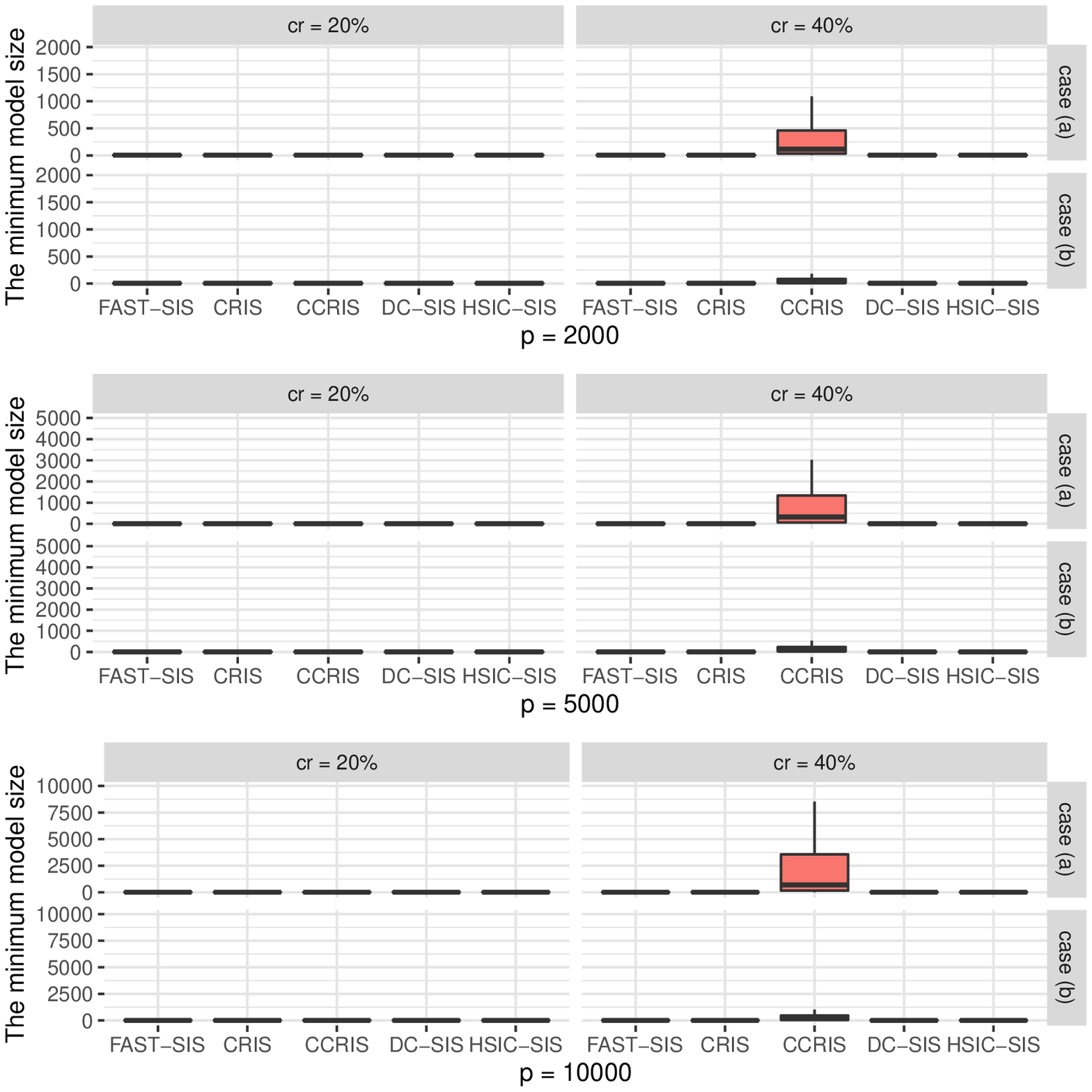}
    \caption{
   Boxplot of the minimum model size that are required to include all active predictors for Example 1 with the censoring rate of 20\% and 40\%. Case (a): random censoring, $ C \sim \rm{Unif}(0,c)$; Case (b): nonrandom censoring,
    $C \sim \rm{Unif}(0, c\cdot\lvert Z_{1}-Z_{2}\rvert)$
    }
	\label{figure1}
\end{figure}

\begin{figure}[htbp]
	\centering
	\includegraphics[width=0.6\textheight]{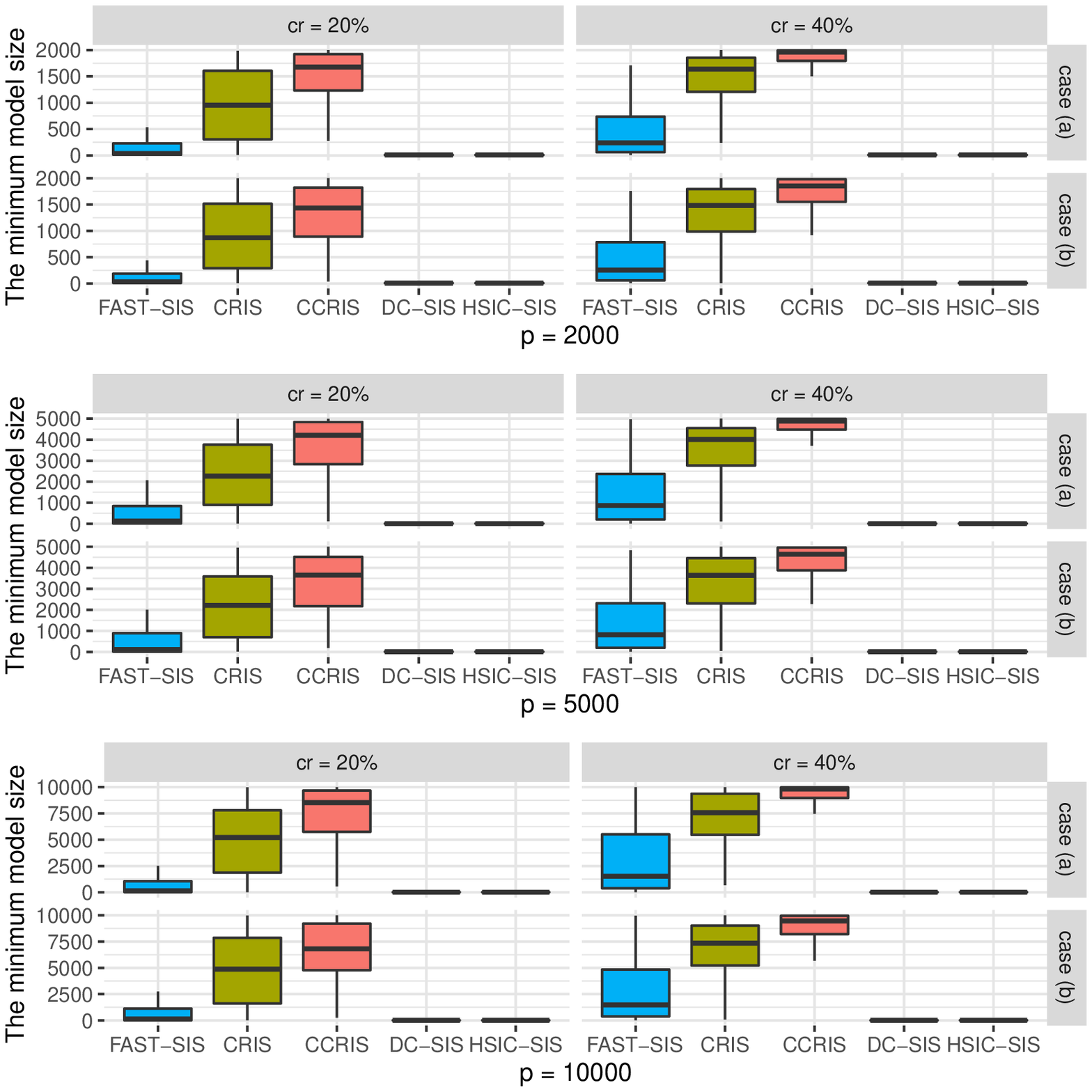}
    \caption{Boxplot of the minimum model size that are required to include all active predictors for Example 2 with the censoring rate of 20\% and 40\%. Case (a): random censoring, $ C \sim \rm{Unif}(0,c)$; Case (b): nonrandom censoring,
    $ C \sim \rm{Unif}(0,c\cdot\lvert Z_{1} - Z_{2}\rvert)$
    }
	\label{figure2}
\end{figure}

\begin{figure}[htbp]
	\centering
	\includegraphics[width=0.6\textheight]{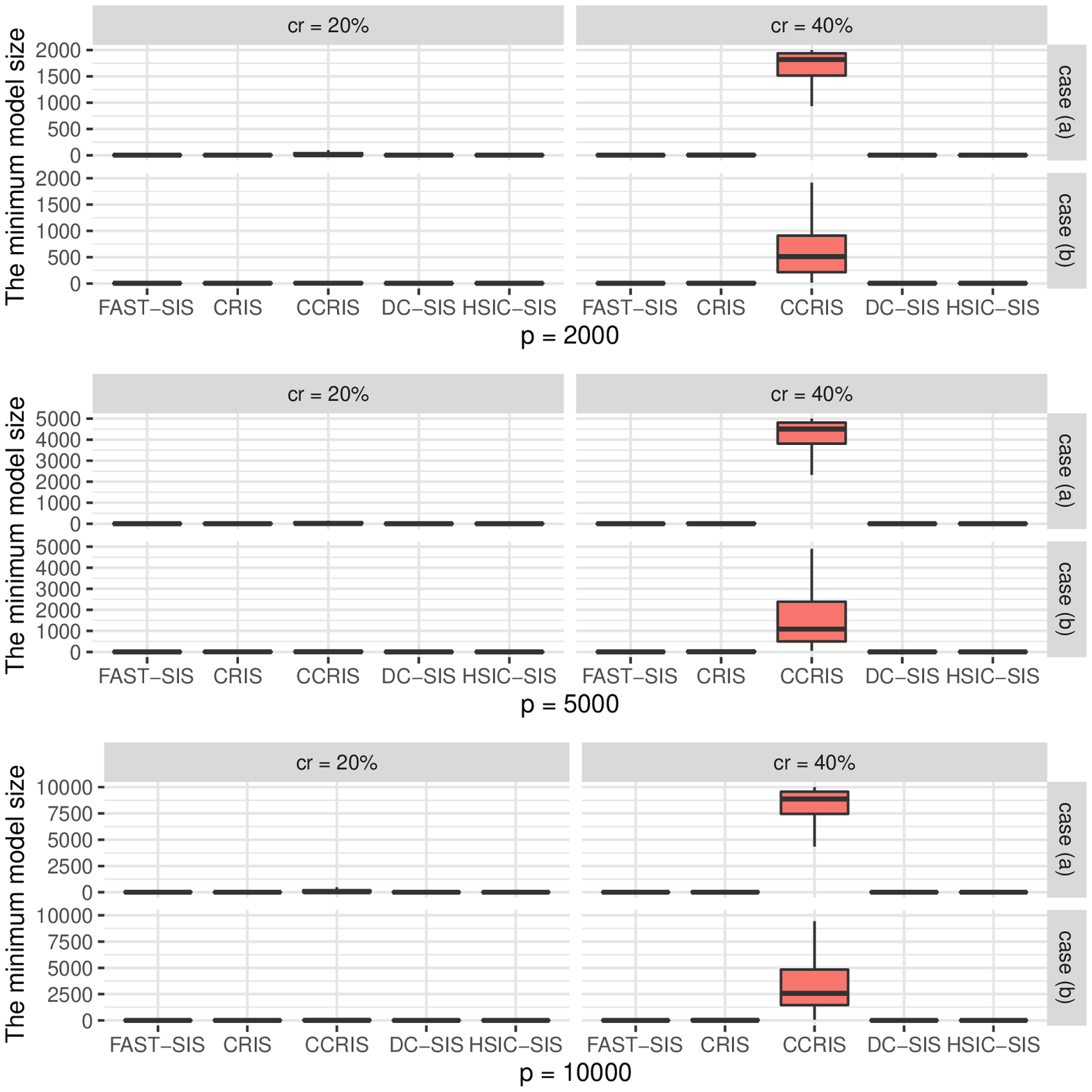}
    \caption{Boxplot of the minimum model size that are required to include all active predictors for Example 3 with the censoring rate of 20\% and 40\%. Case (a): random censoring, $ C \sim \rm{Unif}(0,c)$; Case (b): nonrandom censoring,
    $ C \sim \rm{Unif}(0,c\cdot\lvert Z_{1} - Z_{2}\rvert)$
    }
	\label{figure3}
\end{figure}

To provide comprehensive insight of  these methods,
we further plot the boxplot of $\cal S$ out of 200 replications
for Examples 1-3 in Figures \ref{figure1}-\ref{figure3}.
We can see clearly that the performance of the proposed
HSIC-SIS procudure is robust for different settings
and exhibits more satisfactory
results than  the other considered methods.
Therefore, it could be expected to have
a good application prospect in ultrahigh-dimensional
survival data analysis.

\section{Real Data Analysis}
As an illustration, we apply the proposed screening procedure HSIC-SIS
to the diffuse large-B-cell lymphoma (DLBCL) data  and the ovarian cancer data.
We first demonstrate the reliability and superiority of the proposed HSIC-SIS method
by comparing different screening results of the well-known DLBCL data,
then apply the HSIC-SIS method to the new ovarian cancer data in an attempt to identify the important RNA which affects the survival risk of ovarian cancer patients.

\subsection{The analysis of DLBCL data}
The diffuse large-B-cell lymphoma (DLBCL) data is a classic data which
has been  analyzed in the literature related to feature screening for
ultrahigh-dimensional survival data.
This dataset was originally collected by \citet{Rosenwald2002},
including $n = 240$ patients whose survival time $T$ ranged from $0$ to $21.8$
years, with a median of $2.8$ years across all samples.
Of the $240$ patients, $102$  were still alive at the final
follow-up visit, which led to the censoring rate of $42.5\%$.
For each patient, there are $p = 7399$ gene expression levels,
and the goal is to find the important genes which have great influence on $T$ from $7399$ genes.

For comparison, we also consider
FAST-SIS of \citet{Gorst2013},
CRIS of  \citet{Song2014},
CCRIS of \citet{Zhang2018}, and
DC-SIS of \citet{Zhang2021}
to analyze this data.
Specifically, we apply these five screening methods
to screening the important ones among the 7399 genes
and  select the top $43 = [240/\log(240)]$ covariates.
The selection result of HSIC-SIS is
largely coincides with FAST-SIS and DC-SIS,
which contains 25 and 37 overlapping genes, respectively.
Among the top 43 genes selected by the proposed HSIC-SIS procedure,
there are 41 genes that are also considered  important by at least one other screening method.
Furthermore, we fit a Cox proportional hazards model
based on the top 43 genes selected  by HSIC-SIS,
and utilize the regularization
methods LASSO, SCAD and MCP to select the significant ones among these 43 covariates,
where the tuning parameter is selected by the 10-fold cross-validation.
The unique identifications (UNIQIDs)
and the estimated value of the coefficient of selected covariates
are summarized in Table \ref{realdata1},
from which we draw similar conclusion as \citet{Zhang2021}.
In particular,
genes 31981, 31669, 32238, 24376 and 25054 are all selected by LASSO, SCAD and MCP
methods,
indicating that these genes could be strongly associated with patients' survival risk.
Moreover, the gene 31981 has the greatest impact on the survival risk.
In contrast, HSIC-SIS method attaches more importance to genes 27731, 28377, 27774, 29912, 24367 and 31806 rather than genes 32679, 28641 and 17902.

\begin{figure}[htbp]
	\centering
	\includegraphics[width=0.6\textheight]{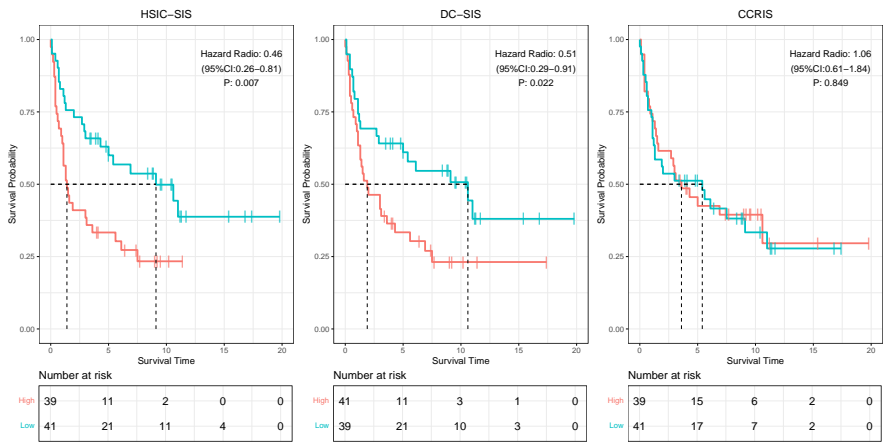}
    \caption{The Kaplan\--Meier estimate of survival curves for the low-risk and high-risk groups from the testing dataset}
	\label{figure.rd1}
\end{figure}

In addition, to evaluate the predictive performance of HSIC-SIS,
240 patients are randomly divided into a training set
with $n_{1} =160$ subjects and a test set with $n_{2} =80$
subjects (\citealp{Rosenwald2002}),
where the training set is used to build the prediction model and the test data is used to evaluate the model.
A summary of the survival times in the DLBCL dataset is shown in Table \ref{realdata2}.
We first apply these five screening methods to the training set
and  select the top $[160/\log(160)]=31$ covariates,
then fit the Cox proportional hazards model on these selected covariates and
perform the regularization method LASSO to further remove some irrelevant covariates,
where the tuning parameters are determined by the 10-fold cross-validation.
Based on the estimated value of the coefficient of selected predictors obtained from the training set,
we calculate  the risk scores for the patients in the testing set
and divide them into low-risk and high-risk groups,
where the threshold is the median of the estimated scores from the training set.
Figure \ref{figure.rd1} depicts the Kaplan--Meier survival curves for each group.
For brevity, we only show the Kaplan--Meier curves for the proposed procedure HSIC-SIS,
and the other two methods  DC-SIS, CCRIS which has the best and the worst separation among the other four screening methods.
We observe that the two curves are well-separated for the proposed HSIC-SIS procedure.
Objectively, we evaluate the difference between the two survival curves by the $\log$-rank test
and summarize the  $p$-values in Figure \ref{figure.rd1},
which shows that HSIC-SIS procedure has the smallest $p$-values,
indicating that HSIC-SIS procedure  behaves a favorable and reliable prediction
based on the final selected model.

\subsection{The analysis of ovarian cancer data}
Ovarian cancer is one of the three major gynecological tumors.
At present, the main treatment method is ovarian cytoreductive surgery plus chemotherapy.
However, secondary drug resistance often occurs.
Studies have shown that genetic factors are closely related to the progression of ovarian cancer disease, while RNA plays a key role in the cell cycle.
Therefore, identifying these RNAs with inducing or inhibitory effects can help target therapy to patients and improve the development strategy of new drugs.
Different researchers have measured RNA expression level data of ovarian cancer patients in different years, which are summarized in the R package ``{\emph{{OvarianCancer}}}".
We explore the GSE51088\_eset dataset, which was determined by \citet{Karlan2014}
by using RNA expression microarray technology.
After deleting the missing values and covariates with variance 0,
we obtain the survival time $T$ and $p = 6844$ RNA expression levels of $n = 152$ patients.
The observed survival time of patients range from 30 to 7001 days, with a median of 1491 days.
Among the 152 patients, 40 were still alive at the last follow-up,
which led to the censoring rate of $26.32\%$.

Similarly, we apply FAST-SIS, CRIS, CCRIS, DC-SIS and HSIC-SIS to screening
the important ones among the 6844 RNAs, and reserve the first $30 = [152/\log(152)]$ RNAs.
Among the 30 covariates selected by HSIC-SIS procedure, there are 22  RNAs  overlapped with DC-SIS and 20 RNAs overlapped with CCRIS,
furthermore,  23 RNAs are also considered as vital covariates by at least one other screening method.

Innovatively, we integrate the results of the five screening procedures using the
voting method.
Voting is an ensemble learning procedure that follows the principle of the minority obeying the majority.
By integrating  multiple models, voting improves the model's robustness and generalization ability.
Specifically, we build a Cox proportional hazards model
using the 25 RNAs with the highest frequency in the five screening results,
and utilize regularization methods LASSO, SCAD and MCP to perform further variable selection
with the tuning parameters selected by the 10-fold cross-validation.
The selection results for significant RNAs and the estimated value of the coefficient of selected predictors are given in Table \ref{realdata3},
from which we can see that
RNA FPR1, ATP2C2, NUDT7, SMIM14 and FAM189A2 are all selected by LASSO, SCAD and MCP,
means that these five RNAs could be strongly associated with patients' survival risk
and the RNA FPR1 may have the greatest impact on the survival risk.
Furthermore,
we consult the literature related to the selected RNA
(\citealp{Ahmet2020,Yan2020,Minopoli2019,Kohn2014,Yang2020,Gendoo2019}),
which shows that our selection results are consistent with theirs.
For example,
\citet{Ahmet2020}, \citet{Yan2020}, \citet{Minopoli2019}
emphasized the importance of RNA FPR1,
where \citet{Minopoli2019} identified FPR1 as a novel and valuable marker for predicting the propensity of ovarian cancer cells to adhere and subsequently invade mesothelium, and suggested FPR1 as a new therapeutic target for the treatment of metastatic ovarian cancer.
\citet{Kohn2014} found that Ca(2+)-related gene ATP2C2 exhibited highly selective expression in epithelial tumor cells.
\citet{Yang2020} listed SMIM14 as a down-regulated expressed gene in ovarian cancer tissues compared to normal controls from four microarray profile datasets.
\citet{Gendoo2019} found that FAM189A2 had significant role in ovarian cancer and was also identified as the only gene that is indicative of worse outcomes.

\section{Conclusion}
In this paper, we proposed a new model-free screening method HSIC-SIS
for ultrahigh dimensional right-censored  survival data
based on the Hilbert-Schmidt independence criterion (HSIC).
It is a model-free procedure and does not rely on
any model structure, thus it works well for various types of survival models.
Unlike most existing model-free screening methods for right-censored  survival data,
this procedure does not involve the Kaplan--Meier estimator,
thus its performance is much more robust to the existence of heavy censoring or
outliers.
Similar to distance correlation,
HSIC is also a measure of independence rather than correlation,
HSIC is zero means that the random variables are independent,
so its performance is better than most existing screening methods,
which can also be seen from the simulation results.
Moreover, the empirical estimate of HSIC is very simple as it is just the trace
of a product of Gram matrices,
thus the corresponding calculation of the proposed procedure is very simple and fast.
The numerical
results indicated that the proposed methodology works well
for practical situations and has its distinctive
advantages for the complicated nonlinear models which
could be more feasible to capture the characteristic
of the ultrahigh-dimensional survival data.

Note that the calculation of
marginal utility $\wh{\omega}_k$ depends on  the selection of kernel function,
since the goal of feature screening is to filter out a majority of inactive variables,
not focus on the accuracy of parameter estimation,
different kernel functions have very little influence on the screening results.
In simulation studies,
we considered the Gaussian kernel with $f(x,y)=\exp(-\frac{\|x-y\|^2}{2\gamma^2})$
and set $\gamma=2$.
We also tried other kernel functions, such as linear kernel function and
Laplacian kernel function, the proposed method also performed well,
we omitted the results here to save space.
Moreover, as the definition of HSIC is very complex,
how to construct the sure screening property and the ranking consistency of the
proposed procedure is a complex question,
we leave it for further research.

There are some issues that deserve further consideration.
First, similar to most marginal screening methods,
the proposed method may fail to detect
the hidden active variables which are jointly important
but are weakly correlated with the response,
how to propose a screening method using the joint information of covariates
is an important issue.
Moreover, if we have prior information on active covariates,
how to include such information to propose more efficient screening methods
merits further investigation.
Finally,
we considered the right-censored survival data in this paper,
while in medical follow-up studies we often collect interval-censored
survival data, how to extend this procedure to handle interval-censored data is
also an interesting problem.

\section*{Acknowledgement}
This work is supported by the
National Natural Science Foundation of China (No: 11971362,11901581),
Natural Science Foundation of Hubei Province (No:2021CFB502).

\section*{Conflict of interest}
The authors declare that we have no conflict of interest.


\begin{table}
\caption{The simulation results for Example 1 with the censoring rate CR$= 20\%$}
\label{ex1.1}
\begin{center}
{\setlength{\tabcolsep}{1.5mm}
\begin{tabular}{lllcccccccccc}
\toprule

&&&&&\multicolumn{5}{c}{${\cal P}_e$\footnotemark[3]}&\\
\cline{6-10}
$p$&Case &Method &Med.\footnotemark[4]&IQR\footnotemark[5]& $X_1$ & $X_2$ & $X_3$  &$X_4$ & $X_5$ &${\cal P}_a$\footnotemark[6]\\
\hline

2000& (a)\footnotemark[1]
 & FAST-SIS& 5& 0& 1.000 & 1.000 & 1.000 & 1.000  & 1.000 & 1.000\\
&& CRIS    & 5& 0& 0.995 & 1.000 & 0.995 & 1.000  & 0.985 & 0.980\\
&& CCRIS   & 5& 0& 1.000 & 1.000 & 1.000 & 1.000  & 1.000 & 1.000\\
&& DC-SIS  & 5& 0& 1.000 & 1.000 & 1.000 & 1.000  & 1.000 & 1.000\\
&& HSIC-SIS& 5& 0& 1.000 & 1.000 & 1.000 & 1.000  & 1.000 & 1.000\\
& (b)\footnotemark[2]
 & FAST-SIS& 5& 0& 1.000 & 1.000 &1.000 & 1.000  & 1.000 & 1.000\\
&& CRIS    & 5& 0& 0.985 & 0.995 &1.000 & 1.000  & 0.980 & 0.965\\
&& CCRIS   & 5& 0& 1.000 & 1.000 &1.000 & 1.000  & 1.000 & 1.000\\
&& DC-SIS  & 5& 0& 1.000 & 1.000 &1.000 & 1.000  & 1.000 & 1.000\\
&&HSIC-SIS & 5& 0& 1.000 & 1.000 &1.000 & 1.000  & 1.000 & 1.000\\
5000& (a)
& FAST-SIS& 5& 0& 1.000 & 1.000 &1.000 & 1.000  & 1.000 & 1.000\\
&& CRIS   & 5& 0& 1.000 & 1.000 &1.000 & 1.000  & 0.985 & 0.985\\
&& CCRIS  & 5& 0& 1.000 & 1.000 &1.000 & 1.000  & 1.000 & 1.000\\
&& DC-SIS & 5& 0& 1.000 & 1.000 &1.000 & 1.000  & 1.000 & 1.000\\
&&HSIC-SIS& 5& 0& 1.000 & 1.000 &1.000 & 1.0000 & 1.000 & 1.000\\
& (b)
& FAST-SIS& 5& 0& 1.000 & 1.000 &1.000 & 1.000  & 1.000 & 1.000\\
&& CRIS   & 5& 1& 1.000 & 1.000 &1.000 & 1.000  & 0.995 & 0.995\\
&& CCRIS  & 5& 0& 1.000 & 1.000 &1.000 & 1.000  & 1.000 & 1.000\\
&& DC-SIS & 5& 0& 1.000 & 1.000 &1.000 & 1.000  & 1.000 & 1.000\\
&&HSIC-SIS& 5& 0& 1.000 & 1.000 &1.000 & 1.000  & 1.000 & 1.000\\
10000& (a)
& FAST-SIS& 5& 0& 1.000 & 1.000 &1.000 & 1.000  & 1.000 & 1.000\\
&& CRIS   & 5& 0& 0.995 & 1.000 &1.000 & 1.000  & 0.990 & 0.985\\
&& CCRIS  & 5& 0& 1.000 & 1.000 &1.000 & 1.000  & 0.995 & 0.995\\
&& DC-SIS & 5& 0& 1.000 & 1.000 &1.000 & 1.000  & 1.000 & 1.000\\
&&HSIC-SIS& 5& 0& 1.000 & 1.000 &1.000 & 1.000  & 1.000 & 1.000\\
& (b)
& FAST-SIS& 5& 0& 1.000 & 1.000 &1.000 & 1.000  & 1.000 & 1.000\\
&& CRIS   & 5& 0& 0.995 & 1.000 &1.000 & 1.000  & 0.990 & 0.985\\
&& CCRIS  & 5& 1& 0.995 & 1.000 &1.000 & 1.000  & 1.000 & 0.995\\
&& DC-SIS & 5& 0& 1.000 & 1.000 &1.000 & 1.000  & 1.000 & 1.000\\
&&HSIC-SIS& 5& 0& 1.000 & 1.000 &1.000 & 1.000  & 1.000 & 1.000\\
\hline
\end{tabular}
}
\end{center}
\vspace{0ex}
\footnotetext[1]{Case (a): random censoring with $C \sim \rm{Unif}(0,c)$;}
\footnotetext[2]{Case (b): nonrandom censoring with $ C \sim \rm{Unif}(0,c\cdot\lvert Z_{1} - Z_{2}\rvert)$;}
\footnotetext[3]{${\cal {P}}_e$: the selection proportion for each active variable;}
\footnotetext[4]{Med.: the median of $\cal S$ over 200 replications;}
\footnotetext[5]{IQR: the interquartile range of $\cal S$;}
\footnotetext[6]{${\cal {P}}_a$: the selection proportion for all active variables.}
\end{table}

\begin{table}
\caption{The simulation results for Example 1 with the censoring rate CR$= 40\%$}
\label{ex1.2}
\begin{center}
{\setlength{\tabcolsep}{1.5mm}
\begin{tabular}{lllrrllllll}
\toprule

&&&&&\multicolumn{5}{c}{${\cal P}_e$\footnotemark[3]}&\\
\cline{6-10}
$p$&Case &Method &Med.\footnotemark[4]&IQR\footnotemark[5]& $X_1$ & $X_2$ & $X_3$  &$X_4$ & $X_5$ &${\cal P}_a$\footnotemark[6]\\
\hline

2000 & (a)\footnotemark[1]
& FAST-SIS  & 5   & 0    & 1.000 & 1.000 & 1.000 & 1.000 & 1.000 & 1.000               \\
&& CRIS     & 5   & 1    & 0.990 & 1.000 & 0.995 & 0.995 & 0.985 & 0.980               \\
&& CCRIS    & 116 & 427  & 0.405 & 0.625 & 0.660 & 0.585 & 0.445 & 0.260               \\
&& DC-SIS   & 5   & 0    & 1.000 & 1.000 & 1.000 & 1.000 & 1.000 & 1.000               \\
&& HSIC-SIS & 5   & 0    & 1.000 & 1.000 & 1.000 & 1.000 & 1.000 & 1.000               \\
& (b)\footnotemark[2]
& FAST-SIS  & 5   & 0    & 1.000 & 1.000 & 1.000 & 1.000 & 1.000 & 1.000               \\
&& CRIS     & 5   & 1    & 0.960 & 0.980 & 0.990 & 0.990 & 0.975 & 0.945               \\
&& CCRIS    & 30  & 73   & 0.670 & 0.945 & 0.965 & 0.925 & 0.785 & 0.545               \\
&& DC-SIS   & 5   & 0    & 1.000 & 1.000 & 1.000 & 1.000 & 1.000 & 1.000               \\
&& HSIC-SIS & 5   & 0    & 1.000 & 1.000 & 1.000 & 1.000 & 1.000 & 1.000               \\
5000 & (a)
& FAST-SIS  & 5   & 0    & 1.000 & 1.000 & 1.000 & 1.000 & 1.000 & 1.000               \\
&& CRIS     & 5   & 1    & 0.990 & 1.000 & 1.000 & 1.000 & 0.970 & 0.965               \\
&& CCRIS    & 324 & 1278 & 0.335 & 0.470 & 0.525 & 0.510 & 0.330 & 0.190               \\
&& DC-SIS   & 5   & 0    & 1.000 & 1.000 & 1.000 & 1.000 & 1.000 & 1.000               \\
&& HSIC-SIS & 5   & 0    & 1.000 & 1.000 & 1.000 & 1.000 & 1.000 & 1.000               \\
& (b)
& FAST-SIS  & 5   & 0    & 1.000 & 1.000 & 1.000 & 1.000 & 1.000 & 1.000               \\
&& CRIS     & 5   & 2    & 0.985 & 0.995 & 1.000 & 1.000 & 0.995 & 0.975               \\
&& CCRIS    & 77  & 206  & 0.455 & 0.825 & 0.890 & 0.835 & 0.615 & 0.325               \\
&& DC-SIS   & 5   & 0    & 1.000 & 1.000 & 1.000 & 1.000 & 0.998 & 0.998               \\
&& HSIC-SIS & 5   & 0    & 1.000 & 1.000 & 1.000 & 1.000 & 1.000 & 1.000               \\
10000 & (a)
& FAST-SIS  & 5   & 0    & 1.000 & 1.000 & 1.000 & 1.000 & 1.000 & 1.000               \\
&& CRIS     & 5   & 1    & 0.980 & 0.990 & 0.995 & 0.995 & 0.990 & 0.980               \\
&& CCRIS    & 698 & 3415 & 0.265 & 0.410 & 0.445 & 0.410 & 0.275 & 0.110               \\
&& DC-SIS   & 5   & 0    & 1.000 & 1.000 & 1.000 & 1.000 & 1.000 & 1.000               \\
&& HSIC-SIS & 5   & 0    & 1.000 & 1.000 & 1.000 & 1.000 & 1.000 & 1.000               \\
& (b)
& FAST-SIS  & 5   & 0    & 1.000 & 1.000 & 1.000 & 1.000 & 1.000 & 1.000               \\
&& CRIS     & 5   & 2    & 0.955 & 0.995 & 1.000 & 0.995 & 0.965 & 0.920               \\
&& CCRIS    & 172 & 425  & 0.330 & 0.745 & 0.845 & 0.780 & 0.560 & 0.230               \\
&& DC-SIS   & 5   & 0    & 1.000 & 1.000 & 1.000 & 1.000 & 1.000 & 1.000               \\
&& HSIC-SIS & 5   & 0    & 1.000 & 1.000 & 1.000 & 1.000 & 1.000 & 1.000               \\
\hline
\end{tabular}
}
\end{center}
\vspace{0ex}
\footnotetext[1]{Case (a): random censoring with $C \sim \rm{Unif}(0,c)$;}
\footnotetext[2]{Case (b): nonrandom censoring with $ C \sim \rm{Unif}(0,c\cdot\lvert Z_{1} - Z_{2}\rvert)$;}
\footnotetext[3]{${\cal {P}}_e$: the selection proportion for each active variable;}
\footnotetext[4]{Med.: the median of $\cal S$ over 200 replications;}
\footnotetext[5]{IQR: the interquartile range of $\cal S$;}
\footnotetext[6]{${\cal {P}}_a$: the selection proportion for all active variables.}
\end{table}

\begin{table}
\caption{The simulation results for Example 2 with the censoring rate CR$= 20\%$}
\label{ex2.1}
\begin{center}
\setlength{\tabcolsep}{2.7mm}
\begin{tabular}{lllrrllll}
\toprule

&&&&&\multicolumn{3}{c}{${\cal P}_e$\footnotemark[3]}&\\
\cline{6-8}
$p$&Case &Method &Med.\footnotemark[4]&IQR\footnotemark[5]& $X_1$ & $X_5$ & $X_{10}$ &${\cal P}_a$\footnotemark[6]\\
\hline

2000 & (a)\footnotemark[1]
& FAST-SIS  & 36   & 215  & 1.000 & 1.000 & 0.515 & 0.515 \\
&& CRIS     & 953  & 1301 & 0.965 & 0.995 & 0.055 & 0.055 \\
&& CCRIS    & 1678 & 691  & 0.935 & 0.640 & 0.000 & 0.000 \\
&& DC-SIS   & 9    & 2    & 1.000 & 1.000 & 0.990 & 0.990 \\
&& HSIC-SIS & 9    & 2    & 1.000 & 1.000 & 1.000 & 1.000 \\
& (b)\footnotemark[2]
& FAST-SIS  & 33   & 174  & 1.000 & 1.000 & 0.520 & 0.520 \\
&& CRIS     & 870  & 1225 & 0.970 & 0.995 & 0.050 & 0.050 \\
&& CCRIS    & 1435 & 930  & 0.820 & 0.575 & 0.010 & 0.000 \\
&& DC-SIS   & 9    & 2    & 1.000 & 1.000 & 0.995 & 0.995 \\
&& HSIC-SIS & 9    & 2    & 1.000 & 1.000 & 1.000 & 1.000 \\
5000 & (a)
& FAST-SIS  & 128  & 826  & 1.000 & 1.000 & 0.335 & 0.335 \\
&& CRIS     & 2268 & 2858 & 0.920 & 0.950 & 0.040 & 0.040 \\
&& CCRIS    & 4204 & 2009 & 0.860 & 0.475 & 0.005 & 0.000 \\
&& DC-SIS   & 10   & 3    & 0.995 & 1.000 & 0.920 & 0.915 \\
&& HSIC-SIS & 10   & 3    & 0.985 & 1.000 & 0.935 & 0.920 \\
& (b)
& FAST-SIS  & 113  & 873  & 1.000 & 1.000 & 0.340 & 0.340 \\
&& CRIS     & 2212 & 2891 & 0.935 & 0.960 & 0.025 & 0.025 \\
&& CCRIS    & 3652 & 2348 & 0.740 & 0.485 & 0.000 & 0.000 \\
&& DC-SIS   & 10   & 4    & 0.995 & 1.000 & 0.915 & 0.910 \\
&& HSIC-SIS & 10   & 3    & 0.985 & 1.000 & 0.925 & 0.915 \\
10000 & (a)
& FAST-SIS  & 140  & 1028 & 1.000 & 1.000 & 0.320 & 0.320 \\
&& CRIS     & 5210 & 5930 & 0.945 & 0.985 & 0.010 & 0.010 \\
&& CCRIS    & 8528 & 3920 & 0.780 & 0.355 & 0.000 & 0.000 \\
&& DC-SIS   & 10   & 2    & 0.995 & 1.000 & 0.970 & 0.965 \\
&& HSIC-SIS & 10   & 2    & 0.995 & 1.000 & 0.970 & 0.965 \\
& (b)
& FAST-SIS  & 134  & 1115 & 1.000 & 1.000 & 0.320 & 0.320 \\
&& CRIS     & 4877 & 6248 & 0.940 & 0.985 & 0.015 & 0.015 \\
&& CCRIS    & 6811 & 4434 & 0.720 & 0.375 & 0.000 & 0.000 \\
&& DC-SIS   & 10   & 3    & 0.995 & 1.000 & 0.975 & 0.970 \\
&& HSIC-SIS & 10   & 3    & 0.995 & 1.000 & 0.980 & 0.975 \\
\hline
\end{tabular}
\end{center}
\vspace{0ex}
\footnotetext[1]{Case (a): random censoring with $C \sim \rm{Unif}(0,c)$;}
\footnotetext[2]{Case (b): nonrandom censoring with $ C \sim \rm{Unif}(0,c\cdot\lvert Z_{1} - Z_{2}\rvert)$;}
\footnotetext[3]{${\cal {P}}_e$: the selection proportion for each active variable;}
\footnotetext[4]{Med.: the median of $\cal S$ over 200 replications;}
\footnotetext[5]{IQR: the interquartile range of $\cal S$;}
\footnotetext[6]{${\cal {P}}_a$: the selection proportion for all active variables.}
\end{table}

\begin{table}
\caption{The simulation results for Example 2 with the censoring rate CR$= 40\%$}
\label{ex2.2}
\begin{center}
\setlength{\tabcolsep}{2.7mm}
\begin{tabular}{lllrrllll}
\toprule

&&&&&\multicolumn{3}{c}{${\cal P}_e$\footnotemark[3]}&\\
\cline{6-8}
$p$&Case &Method &Med.\footnotemark[4]&IQR\footnotemark[5]& $X_1$ & $X_5$ & $X_{10}$ &${\cal P}_a$\footnotemark[6]\\
\hline

2000 & (a)\footnotemark[1]
& FAST-SIS  & 239  & 675  & 1.000 & 1.000 & 0.175 & 0.175 \\
&& CRIS     & 1639 & 646  & 0.580 & 0.640 & 0.005 & 0.005 \\
&& CCRIS    & 1961 & 199  & 0.200 & 0.000 & 0.000 & 0.000 \\
&& DC-SIS   & 9    & 2    & 1.000 & 1.000 & 1.000 & 1.000 \\
&& HSIC-SIS & 9    & 2    & 1.000 & 1.000 & 1.000 & 1.000 \\
& (b)\footnotemark[2]
& FAST-SIS  & 256  & 725  & 1.000 & 1.000 & 0.170 & 0.170 \\
&& CRIS     & 1483 & 806  & 0.680 & 0.745 & 0.015 & 0.015 \\
&& CCRIS    & 1853 & 428  & 0.165 & 0.000 & 0.000 & 0.000 \\
&& DC-SIS   & 9    & 2    & 1.000 & 1.000 & 1.000 & 1.000 \\
&& HSIC-SIS & 9    & 2    & 1.000 & 1.000 & 1.000 & 1.000 \\
5000 & (a)
& FAST-SIS  & 870  & 2165 & 1.000 & 1.000 & 0.125 & 0.125 \\
&& CRIS     & 4008 & 1778 & 0.545 & 0.610 & 0.000 & 0.000 \\
&& CCRIS    & 4884 & 514  & 0.095 & 0.000 & 0.000 & 0.000 \\
&& DC-SIS   & 9    & 2    & 1.000 & 1.000 & 0.960 & 0.960 \\
&& HSIC-SIS & 10   & 4    & 1.000 & 1.000 & 0.935 & 0.935 \\
& (b)
& FAST-SIS  & 816  & 2113 & 1.000 & 1.000 & 0.110 & 0.110 \\
&& CRIS     & 3638 & 2153 & 0.645 & 0.745 & 0.000 & 0.000 \\
&& CCRIS    & 4644 & 1082 & 0.110 & 0.000 & 0.000 & 0.000 \\
&& DC-SIS   & 9    & 3    & 1.000 & 1.000 & 0.970 & 0.970 \\
&& HSIC-SIS & 10   & 4    & 1.000 & 1.000 & 0.945 & 0.945 \\
10000 & (a)
& FAST-SIS  & 1527 & 5134 & 1.000 & 1.000 & 0.075 & 0.075 \\
&& CRIS     & 7579 & 3897 & 0.440 & 0.510 & 0.000 & 0.000 \\
&& CCRIS    & 9830 & 1006 & 0.045 & 0.000 & 0.000 & 0.000 \\
&& DC-SIS   & 9    & 4    & 1.000 & 1.000 & 0.950 & 0.950 \\
&& HSIC-SIS & 10   & 7    & 1.000 & 1.000 & 0.905 & 0.905 \\
& (b)
& FAST-SIS  & 1486 & 4458 & 1.000 & 1.000 & 0.060 & 0.060 \\
&& CRIS     & 7348 & 3790 & 0.570 & 0.625 & 0.000 & 0.000 \\
&& CCRIS    & 9466 & 1744 & 0.070 & 0.000 & 0.000 & 0.000 \\
&& DC-SIS   & 9    & 4    & 1.000 & 1.000 & 0.960 & 0.900 \\
&& HSIC-SIS & 10   & 6    & 1.000 & 1.000 & 0.920 & 0.920 \\
\hline
\end{tabular}
\end{center}
\vspace{0ex}
\footnotetext[1]{Case (a): random censoring with $C \sim \rm{Unif}(0,c)$;}
\footnotetext[2]{Case (b): nonrandom censoring with $ C \sim \rm{Unif}(0,c\cdot\lvert Z_{1} - Z_{2}\rvert)$;}
\footnotetext[3]{${\cal {P}}_e$: the selection proportion for each active variable;}
\footnotetext[4]{Med.: the median of $\cal S$ over 200 replications;}
\footnotetext[5]{IQR: the interquartile range of $\cal S$;}
\footnotetext[6]{${\cal {P}}_a$: the selection proportion for all active variables.}
\end{table}

\begin{table}
\caption{The simulation results for Example 3 with the censoring rate CR$= 20\%$}
\label{ex3.1}
\begin{center}
{\setlength{\tabcolsep}{2.2mm}
\begin{tabular}{lllrrlllll}
\toprule

&&&&&\multicolumn{4}{c}{${\cal P}_e$\footnotemark[3]}&\\
\cline{6-9}
$p$&Case &Method &Med.\footnotemark[4]&IQR\footnotemark[5]& $X_1$ & $X_2$ & $X_9$ & $X_{10}$ &${\cal P}_a$\footnotemark[6]\\
\hline

2000 & (a)\footnotemark[1]
& FAST-SIS  & 4  & 0  & 1.000 & 1.000 & 1.000 & 1.000 & 1.000 \\
&& CRIS     & 4  & 1  & 0.985 & 0.965 & 0.965 & 1.000 & 0.930 \\
&& CCRIS    & 10 & 37 & 0.965 & 0.945 & 0.810 & 0.895 & 0.725 \\
&& DC-SIS   & 4  & 0  & 1.000 & 1.000 & 1.000 & 1.000 & 1.000 \\
&& HSIC-SIS & 4  & 0  & 1.000 & 1.000 & 1.000 & 1.000 & 1.000 \\
& (b)\footnotemark[2]
& FAST-SIS  & 4  & 0  & 1.000 & 1.000 & 1.000 & 1.000 & 1.000 \\
&& CRIS     & 4  & 3  & 0.980 & 0.955 & 0.980 & 1.000 & 0.935 \\
&& CCRIS    & 6  & 10 & 0.985 & 0.955 & 0.940 & 0.985 & 0.885 \\
&& DC-SIS   & 4  & 0  & 1.000 & 1.000 & 1.000 & 1.000 & 1.000 \\
&& HSIC-SIS & 4  & 0  & 1.000 & 1.000 & 1.000 & 1.000 & 1.000 \\
5000 & (a)
& FAST-SIS  & 4  & 0  & 1.000 & 1.000 & 1.000 & 1.000 & 1.000 \\
&& CRIS     & 4  & 6  & 0.965 & 0.950 & 0.950 & 0.995 & 0.880 \\
&& CCRIS    & 14 & 61 & 0.950 & 0.945 & 0.815 & 0.870 & 0.690 \\
&& DC-SIS   & 4  & 0  & 1.000 & 1.000 & 1.000 & 1.000 & 1.000 \\
&& HSIC-SIS & 4  & 0  & 1.000 & 1.000 & 1.000 & 1.000 & 1.000 \\
& (b)
& FAST-SIS  & 4  & 0   & 1.000 & 1.000 & 1.000 & 1.000 & 1.000 \\
&& CRIS     & 5  & 7   & 0.960 & 0.935 & 0.925 & 0.995 & 0.850 \\
&& CCRIS    & 5  & 7   & 0.960 & 0.935 & 0.925 & 0.995 & 0.850 \\
&& DC-SIS   & 4  & 0   & 1.000 & 1.000 & 1.000 & 1.000 & 1.000 \\
&& HSIC-SIS & 4  & 0   & 1.000 & 1.000 & 1.000 & 1.000 & 1.000 \\
10000 & (a)
& FAST-SIS  & 4  & 0   & 1.000 & 1.000 & 1.000 & 1.000 & 1.000 \\
&& CRIS     & 5  & 8   & 0.940 & 0.930 & 0.955 & 0.965 & 0.835 \\
&& CCRIS    & 42 & 188 & 0.885 & 0.885 & 0.695 & 0.790 & 0.490 \\
&& DC-SIS   & 4  & 0   & 1.000 & 1.000 & 1.000 & 1.000 & 1.000 \\
&& HSIC-SIS & 4  & 0   & 1.000 & 1.000 & 1.000 & 1.000 & 1.000 \\
& (b)
& FAST-SIS  & 4  & 0  & 1.000 & 1.000 & 1.000 & 1.000 & 1.000 \\
&& CRIS     & 5  & 10 & 0.920 & 0.925 & 0.950 & 0.975 & 0.830 \\
&& CCRIS    & 10 & 37 & 0.930 & 0.885 & 0.900 & 0.975 & 0.720 \\
&& DC-SIS   & 4  & 0  & 1.000 & 1.000 & 1.000 & 1.000 & 1.000 \\
&& HSIC-SIS & 4  & 0  & 1.000 & 1.000 & 1.000 & 1.000 & 1.000 \\
\hline
\end{tabular}
}
\end{center}
\vspace{0ex}
\footnotetext[1]{Case (a): random censoring with $C \sim \rm{Unif}(0,c)$;}
\footnotetext[2]{Case (b): nonrandom censoring with $ C \sim \rm{Unif}(0,c\cdot\lvert Z_{1} - Z_{2}\rvert)$;}
\footnotetext[3]{${\cal {P}}_e$: the selection proportion for each active variable;}
\footnotetext[4]{Med.: the median of $\cal S$ over 200 replications;}
\footnotetext[5]{IQR: the interquartile range of $\cal S$;}
\footnotetext[6]{${\cal {P}}_a$: the selection proportion for all active variables.}\end{table}

\begin{table}
\caption{The simulation results for Example 3 with the censoring rate CR$= 40\%$}
\label{ex3.2}
\begin{center}
{\setlength{\tabcolsep}{2.2mm}
\begin{tabular}{lllrrlllll}
\toprule

&&&&&\multicolumn{4}{c}{${\cal P}_e$\footnotemark[3]}&\\
\cline{6-9}
$p$&Case&Method &Med.\footnotemark[4]&IQR\footnotemark[5]& $X_1$ & $X_2$ & $X_9$ & $X_{10}$ &${\cal P}_a\footnotemark[6]$\\
\hline

2000 & (a)\footnotemark[1]
& FAST-SIS  & 4    & 0    & 1.000 & 1.000 & 1.000 & 1.000 & 1.000 \\
&& CRIS     & 5    & 7    & 0.950 & 0.910 & 0.975 & 0.995 & 0.855 \\
&& CCRIS    & 1820 & 423  & 0.125 & 0.110 & 0.005 & 0.005 & 0.000 \\
&& DC-SIS   & 4    & 0    & 1.000 & 1.000 & 1.000 & 1.000 & 1.000 \\
&& HSIC-SIS & 4    & 0    & 1.000 & 1.000 & 1.000 & 1.000 & 1.000 \\
& (b)\footnotemark[2]
& FAST-SIS  & 4    & 0    & 1.000 & 1.000 & 1.000 & 1.000 & 1.000 \\
&& CRIS     & 5    & 12   & 0.925 & 0.930 & 0.940 & 0.985 & 0.840 \\
&& CCRIS    & 511  & 693  & 0.400 & 0.350 & 0.360 & 0.385 & 0.020 \\
&& DC-SIS   & 4    & 0    & 1.000 & 1.000 & 1.000 & 1.000 & 1.000 \\
&& HSIC-SIS & 4    & 0    & 1.000 & 1.000 & 1.000 & 1.000 & 1.000 \\
5000 & (a)
& FAST-SIS  & 4    & 0    & 1.000 & 1.000 & 1.000 & 1.000 & 1.000 \\
&& CRIS     & 5    & 24   & 0.880 & 0.860 & 0.935 & 1.000 & 0.770 \\
&& CCRIS    & 4500 & 1000 & 0.045 & 0.085 & 0.010 & 0.000 & 0.000 \\
&& DC-SIS   & 4    & 0    & 1.000 & 1.000 & 1.000 & 1.000 & 1.000 \\
&& HSIC-SIS & 4    & 0    & 1.000 & 1.000 & 1.000 & 1.000 & 1.000 \\
& (b)
& FAST-SIS  & 4    & 0    & 1.000 & 1.000 & 1.000 & 1.000 & 1.000 \\
&& CRIS     & 8    & 33   & 0.900 & 0.900 & 0.910 & 0.985 & 0.755 \\
&& CCRIS    & 1087 & 1889 & 0.260 & 0.195 & 0.275 & 0.280 & 0.000 \\
&& DC-SIS   & 4    & 0    & 1.000 & 1.000 & 1.000 & 1.000 & 1.000 \\
&& HSIC-SIS & 4    & 0    & 1.000 & 1.000 & 1.000 & 1.000 & 1.000 \\
10000 & (a)
& FAST-SIS  & 4    & 0    & 1.000 & 1.000 & 1.000 & 1.000 & 1.000 \\
&& CRIS     & 7    & 44   & 0.835 & 0.835 & 0.955 & 0.980 & 0.725 \\
&& CCRIS    & 8872 & 2109 & 0.035 & 0.020 & 0.000 & 0.000 & 0.000 \\
&& DC-SIS   & 4    & 0    & 1.000 & 1.000 & 1.000 & 1.000 & 1.000 \\
&& HSIC-SIS & 4    & 0    & 1.000 & 1.000 & 1.000 & 1.000 & 1.000 \\
& (b)
& FAST-SIS  & 4    & 0    & 1.000 & 1.000 & 1.000 & 1.000 & 1.000 \\
&& CRIS     & 11   & 53   & 0.875 & 0.870 & 0.890 & 0.965 & 0.705 \\
&& CCRIS    & 2572 & 3376 & 0.175 & 0.175 & 0.150 & 0.205 & 0.000 \\
&& DC-SIS   & 4    & 0    & 1.000 & 1.000 & 1.000 & 1.000 & 1.000 \\
&& HSIC-SIS & 4    & 0    & 1.000 & 1.000 & 0.995 & 1.000 & 0.995 \\
\hline
\end{tabular}
}
\end{center}
\vspace{0ex}
\footnotetext[1]{Case (a): random censoring with $C \sim \rm{Unif}(0,c)$;}
\footnotetext[2]{Case (b): nonrandom censoring with $ C \sim \rm{Unif}(0,c\cdot\lvert Z_{1} - Z_{2}\rvert)$;}
\footnotetext[3]{${\cal {P}}_e$: the selection proportion for each active variable;}
\footnotetext[4]{Med.: the median of $\cal S$ over 200 replications;}
\footnotetext[5]{IQR: the interquartile range of $\cal S$;}
\footnotetext[6]{${\cal {P}}_a$: the selection proportion for all active variables.}
\end{table}

\begin{table}
\caption{The results of selected important genes for the diffuse large-B-cell lymphoma data using the regularization methods based on the HSIC-SIS procedure}
\label{realdata1}
\vspace{0ex}
\begin{center}
{\setlength{\tabcolsep}{3.5mm}
\begin{tabular}{crccrccrc}
\hline
\multicolumn{2}{c}{LASSO}&&\multicolumn{2}{c}{SCAD}&&\multicolumn{2}{c}{MCP}\\
\cline{1-2}\cline{4-5}\cline{7-8}
UNIQID\footnotemark[1]&EST.\footnotemark[2]&&UNIQID&EST.&&UNIQID&EST.\\
\hline

31981 &  0.280 & &31981 &  0.400 && 31981 &  0.401 \\
27731 & -0.238 & &27731 & -0.348 && 32238 &  0.351 \\
32238 &  0.227 & &32238 &  0.344 && 27731 & -0.341 \\
24376 & -0.150 & &24376 & -0.220 && 24376 & -0.233 \\
28377 & -0.150 & &28377 & -0.218 && 28377 & -0.228 \\
25054 &  0.146 & &25054 &  0.203 && 25054 &  0.218 \\
31669 &  0.134 & &31669 &  0.160 && 27774 & -0.184 \\
27774 & -0.108 & &27774 & -0.154 && 31669 &  0.183 \\
31242 &  0.073 & &29912 & -0.053 && 29912 & -0.059 \\
17154 & -0.066 & &24367 &  0.043 && 24367 &  0.032 \\
31806 &  0.055 & &31806 &  0.029 && 31806 &  0.013 \\
29912 & -0.052 & &31242 &  0.020 &&       &        \\
25116 &  0.033 & &25116 &  0.019 &&       &        \\
24203 &  0.033 & &17154 & -0.010 &&       &        \\
24367 &  0.032 & &17646 &  0.004 &&       &        \\
23970 &  0.031 & &      &        &&       &        \\
27267 & -0.026 & &      &        &&       &        \\
27592 & -0.024 & &      &        &&       &        \\
28532 & -0.018 & &      &        &&       &        \\
17646 &  0.017 & &      &        &&       &        \\
30669 &  0.011 & &      &        &&       &        \\
33014 &  0.007 & &      &        &&       &        \\

\hline

\end{tabular}
}
\end{center}
\footnotetext[1]{UNIQID: the unique identification;}
\footnotetext[2]{EST.: the estimated value of the coefficient.}
\end{table}

\begin{table}
\caption{Summary of survival times in the diffuse large-B-cell lymphoma (DLBCL) dataset}
\label{realdata2}
\begin{center}
{\setlength{\tabcolsep}{5.3mm}
\begin{tabular}{lrccrr}

\hline

\hline

\hline
Dataset&Num\footnotemark[4]&Min\footnotemark[5]&Max\footnotemark[6]&Median\footnotemark[7]&Cen\footnotemark[8]($\%$)\\
\hline

Train\footnotemark[1]&160&0&21.8&2.5&$45$\\
Test\footnotemark[2]&80&0&19.8&3.45&$37.5$\\
All\footnotemark[3]&240&0&21.8&2.8&$42.5$\\
\hline
\end{tabular}
}
\end{center}
\vspace{0ex}
\footnotetext[1]{Train: the training dataset;}
\footnotetext[2]{Test: the testing dataset;}
\footnotetext[3]{All: the total of training and testing dataset;}
\footnotetext[4]{Num: the number of patients;}
\footnotetext[5]{Min: the minimum observed survival time;}
\footnotetext[6]{Max: the  maximum observed survival time;}
\footnotetext[7]{Median: the median of observed survival time;}
\footnotetext[8]{Cen: the censoring rate.}
\end{table}

%
%
%
%
%

\begin{table}
\caption{The results of selected important RNAs for the ovarian data using the regularization methods based on voting}
\label{realdata3}
\vspace{0ex}
\begin{center}
{\setlength{\tabcolsep}{3mm}
\begin{tabular}{crccrccrc}

\hline
\multicolumn{2}{c}{LASSO}&&\multicolumn{2}{c}{SCAD}&&\multicolumn{2}{c}{MCP}\\
\cline{1-2}\cline{4-5}\cline{7-8}
UNIQID\footnotemark[1]&EST.\footnotemark[2]&&UNIQID&EST.&&UNIQID&EST.\\
\hline

FPR1     &  0.158 && FPR1     &  0.143 && FPR1     &  0.257 \\
NUDT7    & -0.140 && NUDT7    & -0.123 && ATP2C2   & -0.224 \\
SMIM14   & -0.127 && FAM189A2 & -0.117 && NUDT7    & -0.187 \\
FAM189A2 & -0.125 && ATP2C2   & -0.114 && SMIM14   & -0.075 \\
ATP2C2   & -0.119 && SMIM14   & -0.105 && FAM189A2 & -0.070 \\
AK7      & -0.079 && AK7      & -0.065 &&          &        \\
ACSM1    & -0.057 && ACSM1    & -0.035 &&          &        \\
LRRC25   &  0.046 && LRRC25   &  0.032 &&          &        \\
PTGIR    &  0.027 && PTGIR    &  0.018 &&          &        \\
RIBC1    & -0.002 &&          &        &&          &        \\

\hline

\end{tabular}
}
\end{center}
\footnotetext[1]{UNIQID: the unique identification;}
\footnotetext[2]{EST.: the estimated value of the coefficient.}
\end{table}

\end{document}